\documentclass[aps,prd,preprint,floatfix,nofootinbib]{revtex4}

\usepackage{graphicx,epsfig}

\newcommand{\lan}{{\langle}}
\newcommand{\ran}{{\rangle}}

\newcommand{\beq}{\begin{equation}}
\newcommand{\eeq}{\end{equation}}
\newcommand{\bea}{\begin{eqnarray}}
\newcommand{\eea}{\end{eqnarray}}
\newcommand{\nn}{\nonumber}

\begin{document}

\preprint{hep-ph/0602175}


\title{Single Top-Quark Production in Flavor-Changing $Z'$ Models}

\renewcommand{\thefootnote}{\fnsymbol{footnote}}

\author{Abdesslam Arhrib$^{1,2}$
\footnote{Email address: aarhrib@ictp.it}, 
Kingman Cheung$^{1,3}$
\footnote{Email address: cheung@phys.nthu.edu.tw}, 
Cheng-Wei Chiang$^{4,5}$
\footnote{Email address: chengwei@phy.ncu.edu.tw}, 
and Tzu-Chiang Yuan$^{3}$
\footnote{Email address: tcyuan@phys.nthu.edu.tw} }
\affiliation{1.National Center for Theoretical
Sciences,~National~Tsing~Hua~University,~Hsinchu,~Taiwan.}
\affiliation{2.Facult\'e des Sciences et Techniques B.P 416 Tangier,
Morocco.}  
\affiliation{3.Department of Physics, National Tsing Hua University,
Hsinchu, Taiwan, R.O.C.}  
\affiliation{4.Department of Physics,
National Central University, Chungli, Taiwan 320, R.O.C.}
\affiliation{5.Institute of Physics, Academia Sinica, Taipei, Taiwan
115, R.O.C.}

\renewcommand{\thefootnote}{\arabic{footnote}}

\date{\today}

\begin{abstract}
  In some models with an extra $U(1)$ gauge boson $Z'$, the gauge couplings of
  the $Z'$ to different generations of fermions may not be universal. Flavor
  mixing in general can be induced at the tree level in the up-type and/or
  down-type quark sector after diagonalizing their mass matrices.  In this
  work, we concentrate on the flavor mixing in the up-type quark sector.  We
  deduce a constraint from $D^0-\overline{D^0}$ mixing.  We study in detail
  single top-quark production via flavor-changing $Z'$ exchange at the LHC and
  the ILC.  We found that for a typical value of $M_{Z'}=1$ TeV, the production
  cross section at the LHC can be of the order of $1$ fb. However, the
  background from the Standard Model single top-quark production makes it
  difficult to detect the flavor-changing $Z'$ signal unless with a decent
  charm tagging method.  On the other hand, at the ILC, the production cross
  section at the resonance energy of $\sqrt{s} \approx M_{Z'}$ can reach a size
  of more than 100 fb.  Even away from the resonance, the cross section at ILC
  is shown to be larger than the threshold of observability of 0.01 fb.
\end{abstract}

\maketitle

\section{Introduction}

Searches for flavor-changing neutral currents (FCNC) have been pursued for many
years.  So far the sizes of FCNC in the $u$-$c$, $b$-$s$, $s$-$d$, and $b$-$d$
sectors are in general agreement with the Standard Model (SM) predictions,
namely, those given by the Cabibbo-Kobayashi-Maskawa (CKM) mechanism.  In the
SM, tree-level FCNC is absent in both gauge and Yukawa interactions.  They can
only arise from loop diagrams, such as penguin and box diagrams, and are
therefore highly suppressed.  Nevertheless, one-loop FCNC processes can be
enhanced by orders of magnitude in some cases due to the presence of new
physics, see Ref.~\cite{rev} for a review.  Tree-level FCNCs via some exotic
gauge bosons are empirically allowed only if these bosons are sufficiently
heavy or their couplings to SM particles are sufficiently small; otherwise,
they would have been ruled out by current data
\cite{nardi,Langacker:2000ju,Barger:2003hg,Barger:2004qc,Barger:2004hn}.

However, the effects of FCNC involving the top-quark are not yet well probed
experimentally, at least not by the present data.  From the existing LEP and
Tevatron data we have only very weak constraints on the $t$-$q$-$Z$ and
$t$-$q$-$\gamma$ FCNC couplings.  These constraints will not be improved any
further until the operation of the Large Hadron Collider (LHC) or perhaps a
future International Linear Collider (ILC) \cite{aguila} is built.  The goal of
the paper is to analyze effects of tree-level FCNC interactions induced by an
additional $Z'$ boson on the up-type quark sector in general.  In particular,
we study the $t \bar c + \bar t c$ production at the LHC and ILC.

Examples of $Z'$ arising from some grand unified theory (GUT) models are
\cite{nardi}:
\begin{eqnarray}
&& Z_\psi \mbox{ occurring in } E_6 \to SO(10) \times U(1)_\psi ~, \nn \\
&& Z_\chi \mbox{ occurring in } SO(10) \to SU(5) \times U(1)_\chi ~, \nn \\
&& Z_\eta \equiv \cos\theta \, Z_\chi - \sin\theta \, Z_\psi ~, \qquad
  \cos\theta=\sqrt{3/8} \;.\nn
\end{eqnarray}
In these examples, the SM fermions together with an additional right-handed
neutrino are placed in the $\textbf{16}$ of $SO(10)$ embedded in the
$\textbf{27}$ of $E_6$.  One expects in such models that the $Z'$ boson will
couple universally to the three generations of fermions and thus the couplings
are diagonal in the flavor space.\footnote{Since the vector- and
  axial-vector-current interactions of $Z'$ always couple to either two
  left-handed fields or two right-handed fields, the unitary rotations of the
  gauge eigenstates to the mass eigenstates will always preserve the
  diagonality of the $Z'$ interactions if the chiral couplings are
  family-universal.}
However, it is possible that exotic quarks like $h$ and $h^c$ in the
$\textbf{27}$ of $E_6$ may have their $U(1)'$ charges different from the
left-handed and right-handed down-type quarks.  In this case, the SM quarks
will mix, leading to in general both $Z$- and $Z'$-mediated FCNCs.  We note
that flavor-changing $Z'$ boson can also arise in certain dynamical symmetry
breaking models \cite{hill}.

In some string models, the three generations of SM fermions are constructed
differently, resulting in family non-universal $Z'$ couplings to fermions in
different generations.  As a first step, we consider the particular case that
the $Z'$ couples with a different strength to the third generation, as
motivated by a particular class of string models \cite{chaudhuri}.  Once we do
a unitary rotation from the interaction basis to mass eigenbasis, tree-level
FCNCs are induced naturally.  Several works have been done regarding the FCNCs
in the down-type quark sector recently
\cite{Barger:2003hg,Barger:2004qc,Barger:2004hn}.  The same can occur in the
up-type quark sector too.  In order to increase the predictive power of our
model, we assume in this paper that the left-handed down-type sector is already
in diagonal form, such that $V_{\rm CKM} = V_{uL}^\dagger$, where $V_{uL}$ is
the left-handed up-type sector unitary rotation matrix.  Since we do not have
much information about both the right-handed up-type and down-type sectors, we
simply assume that their $Z'$ interactions are family-universal and
flavor-diagonal in the interaction basis.  In this case, unitary rotations keep
the right-handed couplings flavor-diagonal.  Therefore, the only FCNCs arise in
the left-handed $t$-$c$-$u$ sector and depend on the CKM matrix elements and
one additional parameter $x$, which denotes the strength of the $Z'$ coupling
to the third generation relative to the first two generations.  Consequently,
if $x$ is an $O(1)$ parameter but not exactly equal to 1, the $t$-$c$-$Z'$ will
produce the largest FCNC effect.

Associated top-charm production at the LHC or ILC in the SM is expected to be
very suppressed \cite{tcsm}. However, the rates enhanced by the presence of new
physics such as SUSY, topcolor-assisted technicolor or extended Higgs sector
\cite{tcbeyondsm} may reach observable rates in some cases, and can then be
used to probe FCNC couplings.  Single top production can also proceed through
the introduction of anomalous couplings: $t$-$q$-$g$, $t$-$q$-$\gamma$ and
$t$-$q$-$Z$ \cite{Han:1998tp} at both hadron and $e^+e^-$ colliders.  Such
model independent analysis are useful in probing the strength of observable
FCNC couplings.  Many detailed studies of the $Z'$ phenomenology have been done
in recent years
\cite{nardi,Langacker:2000ju,Barger:2003hg,Barger:2004qc,Barger:2004hn,%
Chivukula:2002ry,Yue:2002ja,Chivukula:2003wj,Yue:2003wd,Larios:2003jq,%
Chen:2006vs}.

In this work, we study the capability of the LHC and the ILC to identify the
$t$-$c$ FCNC effect by measuring the production of $t\bar c + \bar t c$ pairs.
Since most of the cross section comes from the $s$-channel production of the
$Z'$, these types of FCNC processes will be searched only after the $Z'$ is
discovered.  The most obvious channel to discover the $Z'$ is the Drell-Yan
process at the LHC, in which a clean resonance peak can be identified in the
invariant mass spectrum $M(\ell^+\ell^-)$ of the lepton-antilepton pair.
Experimenters can then search for the hadronic modes with an invariant mass
reconstructed at the $Z'$ mass.  Those involving the top-quark may be somewhat
complicated because of the 3-jet or 1-jet-1-lepton-$\not\!E_T$ decay products
of the parent top-quark.  But in principle they can be measured, though at
lower efficiencies.  At the LHC, however, the SM single top-quark production
presents a challenging background to $t\bar c + \bar t c$ production.  Unless
one can efficiently distinguish the charm-quark from the bottom-quark and the
other light quarks, the SM single top-quark background makes the FCNC $t\bar c
+ \bar t c$ process very pessimistic.  There may be a slight possibility of
$D$-tagging but it is still too early to tell its efficiency.  On the other
hand, an $e^+ e^-$ collider or the ILC is an ideal place to search for $t\bar c
+ \bar t c$ FCNC production.  One can measure the ratio of the production rates
for $t\bar t: t\bar c+\bar t c : c\bar c$ to identify the FCNC in $t$-$c$
sector.  Also, charm tagging is considerably easier in the $e^+ e^-$
environment.  At any rate, one can simply measure the $t\bar t$ pairs and a
single top-quark plus one jet (either $c$ or $u$) in the hadronic decays of the
$Z'$ boson.  We will estimate the potential of this approach in this paper.

The organization of the paper is as follows.  In the next section, we outline
the formalism of the model.  In Sec.~III, we derive the current limit on the
$c$-$u$ transition from the $D^0$--$\overline{D^0}$ mixing.  We calculate the
production rates of various channels and estimate their detectabilities in
Sec.~IV.  Our conclusion is presented in Sec.~V.

\section{Formalism}

We follow closely the formalism in Ref.\cite{Langacker:2000ju}.  In the gauge
eigenstate basis, the neutral current Lagrangian can be written as
\beq
{\mathcal L}_{\rm NC} = -eJ^{\mu}_{\mbox{\tiny em}} A_{\mu} - g_1
J^{(1)\,\mu} Z^{0}_{1\mu} - g_2 J^{(2)\,\mu} Z^{0}_{2\mu}\,,
\eeq
where $Z^0_1$ is the $SU(2)\times U(1)$ neutral gauge boson, $Z^0_2$ the new
gauge boson associated with an additional Abelian gauge symmetry.  We assume
for simplicity that there is no mixing between $Z^0_1$ and $Z^0_2$, then they
are also the mass eigenstates $Z$ and $Z'$ respectively. The current associated
with the additional $U(1)'$ gauge symmetry is
\beq
  J^{(2)}_{\mu} = \sum\limits_{i,j} \overline{\psi}_i \gamma_{\mu} 
    \left[ \epsilon^{(2)}_{\psi_{L_{ij}}}P_L 
     + \epsilon^{(2)}_{\psi_{R_{ij}}}P_R\right]\psi_j\,,
\eeq
where $\epsilon^{(2)}_{\psi_{{L,R}_{ij}}}$ is the chiral coupling of $Z_2^0$
with fermions $i$ and $j$ running over all quarks and leptons.  If the $Z_2^0$
couplings are diagonal but family-nonuniversal, flavor changing couplings are
induced by fermion mixing.

$Z'$-mediated FCNCs have been studied in detail in Ref.\cite{Barger:2003hg} for
the down-type quark sector and their implications in $B$ meson decays.  Since
such an effect may occur to the up-type quarks as well, we concentrate on this
sector in this paper.  For simplicity, we assume that the $Z'$ couplings to the
leptons and down-type quarks are flavor-diagonal and family-universal:
$\epsilon_{L,R}^d = Q_{L,R}^d {\bf 1}$, $\epsilon_{L,R}^e = Q_{L,R}^e {\bf 1}$
and $\epsilon_{L}^\nu = Q_{L}^\nu {\bf 1}$ where ${\bf 1}$ is the $3 \times 3$
identity matrix in the generation space and $Q_{L,R}^{d}$, $Q_{L,R}^{e}$ and
$Q_{L}^{\nu}$ are the chiral charges.  On the other hand, the interaction
Lagrangian of $Z'$ with the up-type quarks is given by
\begin{equation}
{\mathcal L}^{(2)}_{\rm NC} =
 - g_2  Z'_\mu \; \left( \bar u, \; \bar c,\; \bar t \right)_I \; \gamma^\mu
  \left( \epsilon^u_L P_L + \epsilon^u_R P_R \right ) \; 
  \left(  \begin{array}{c}
                u \\
                c \\
                t \end{array} \right )_I
\end{equation}
where the subscript $I$ denotes the interaction basis.  For definiteness in our
predictions, we assume
\begin{equation}
  \epsilon_L^u = Q^u_L \left( \begin{array}{ccc}
                       1 & 0 & 0 \\
                       0 & 1 & 0 \\
                       0 & 0 & x  \end{array} \right ) \; \; \qquad {\rm and}
                   \qquad \; \; 
  \epsilon_R^u = Q^u_R \left( \begin{array}{ccc}
                       1 & 0 & 0 \\
                       0 & 1 & 0 \\
                       0 & 0 & 1  \end{array} \right ) ~.   \label{epsilonLR}
\end{equation}
That is, only the left-handed couplings are family non-universal.  The
deviation from family universality and thus the magnitude of FCNC are
characterized by the parameter $x$ in the $\overline{t_L}$-$t_L$-$Z'$ entry,
which we take to be of $O(1)$ but not equal to $1$.  $Q^{u}_{L,R}$ are the
chiral $U(1)'$ charges of the up-type quarks. The chiral charges need to be
specified by the $Z'$ model of interest.

When diagonalizing the up-type Yukawa coupling or the mass matrix, we rotate
the left-handed and right-handed fields by $V_{uL}$ and $V_{uR}$, respectively.
Therefore, the Lagrangian ${\mathcal L}^{(2)}_{\rm NC}$ becomes
\begin{equation}
{\mathcal L}^{(2)}_{\rm NC} =
 - g_2  Z'_\mu \; \left( \bar u, \; \bar c,\; \bar t \right)_M \; \gamma^\mu
  \left( V^\dagger_{uL}\epsilon^u_L V_{uL} P_L + V^\dagger_{uR} \epsilon^u_R 
   V_{uR} P_R \right ) \; 
  \left(  \begin{array}{c}
                u \\
                c \\
                t \end{array} \right )_M
\end{equation}
where the subscript $M$ denotes the mass eigenbasis.  With the form of
$\epsilon^u_R$ assumed in Eq.(\ref{epsilonLR}), the right-handed sector is
still flavor-diagonal in the mass eigenbasis, because $\epsilon^u_R$ is
proportional to the identity matrix.  However, $V^\dagger_{uL}\epsilon^u_L
V_{uL}$ is in general non-diagonal.  With the fact that $V_{\rm CKM} =
V_{uL}^\dagger V_{dL}$ and our assumption of the down-quark sector has no
mixing,
\[
 V_{\rm CKM } = V_{uL}^\dagger ~.
\]
The flavor mixing in the left-handed fields is in this case simply related to
$V_{\rm CKM}$, making the model more predictive.  Explicitly,
\begin{eqnarray}
B^{u}_L
&\equiv& V^\dagger_{uL}\epsilon^u_L V_{uL}
= V_{\rm CKM}\epsilon^u_L V^\dagger_{\rm CKM}
\nn \\
& \approx & Q_L^u \left(\begin{array}{ccc}
1  & (x-1) V_{ub} V_{cb}^* & (x-1) V_{ub}V_{tb}^* \\
(x-1) V_{cb} V_{ub}^* & 1  & (x-1) V_{cb} V_{tb}^* \\
(x-1) V_{tb} V_{ub}^* & (x-1) V_{tb} V_{cb}^* & x \\
\end{array}\right) \label{zput}
\end{eqnarray}
where we have used the unitarity conditions of $V_{\rm CKM}$.  It is easy to
see that the sizes of the flavor-changing couplings satisfy in the following
hierarchy: $|B^{tc}_L| > |B^{tu}_L| > |B^{cu}_L|$.  Note that the right-handed
couplings are flavor-diagonal and are of $O(1)$.

The following $Z'$ models will be considered in this work: (i) $Z'$ of the
sequential $Z$ model, (ii) $Z_{LR}$ of the left-right symmetric model, (iii)
$Z_\chi$ occurring in $SO(10) \to SU(5) \times U(1)$, (iv) $Z_\psi$ occurring
in $ E_6 \to SO(10) \times U(1)$, (v) $Z_\eta \equiv \sqrt{3/8} Z_\chi -
\sqrt{5/8} Z_\psi$ occurring in many superstring-inspired models in which $E_6$
breaks directly down to a rank-5 group \cite{luo}.  In the sequential $Z$
model, the gauge coupling $g_2 = g_1$ and the chiral couplings are the same as
the SM $Z$ boson.  In the other models, the gauge coupling takes on the value
\[
g_2 = \sqrt{ \frac{5}{3} }\, \sin \theta_{\rm w} \, g_1 \, \lambda_g^{1/2},
\]
where $\lambda_g$ is $O(1)$ in string-inspired models and $\theta_{\rm w}$ is
the Weinberg angle.  We simply choose $\lambda_g=1$ throughout.  The chiral
couplings of the $Z_{LR}$ in the left-right symmetric model is given by
\cite{luo}
\begin{eqnarray}
Q_L^i &=& - \sqrt{\frac{3}{5}} \left(\frac{1}{2 \alpha} 
  \right )\,(B - L)_i \,, \\
Q_R^i &=& \sqrt{\frac{3}{5}} \left( \alpha T_{3R}^i - 
  \frac{1}{2 \alpha} (B - L)_i \right ) \;,
\end{eqnarray}
where $B$ and $L$ denote the baryon and lepton numbers of the fermion $i$,
respectively.  $T_{3R}$ is the third component of its right-handed isospin in
the $SU(2)_R$ group.  In the left-right symmetric model with $g_L=g_R$, the
parameter $\alpha$ is given by
\[
\alpha = \left( \frac{1 - 2 \sin^2 \theta_{\rm w}}
                     { \sin^2\theta_{\rm w} } \right )^{1/2} \simeq 1.52 ~,
\]
where we have used $\sin^2 \theta_{\rm w} = 0.2316$.  The chiral charges for
these various $Z'$ models are compiled in Table~\ref{table1}.

\begin{table}[b!]
\caption{\small \label{table1}
Chiral couplings of various $Z'$ models.}
\medskip
\begin{ruledtabular}
\begin{tabular}{cccccc}
      & Sequential $Z$ & $Z_{LR}$ & $Z_\chi$ & $Z_\psi$ & $Z_\eta$ \\
\hline
$Q_L^u$ & $0.3456$ & $-0.08493$ &  $\frac{-1}{2\sqrt{10}}$ & 
      $\frac{1}{\sqrt{24}}$ & $\frac{-2}{2\sqrt{15}}$ \\
$Q_R^u$ & $-0.1544$  & $0.5038$ &  $\frac{1}{2\sqrt{10}}$ & 
      $\frac{-1}{\sqrt{24}}$ & $\frac{2}{2\sqrt{15}}$ \\
$Q_L^d$ & $-0.4228$ & $-0.08493$ &  $\frac{-1}{2\sqrt{10}}$ & 
      $\frac{1}{\sqrt{24}}$ & $\frac{-2}{2\sqrt{15}}$ \\
$Q_R^d$ & $0.0772$ & $-0.6736$ &  $\frac{-3}{2\sqrt{10}}$ & 
      $\frac{-1}{\sqrt{24}}$ & $\frac{-1}{2\sqrt{15}}$ \\
$Q_L^e$ & $-0.2684$ & $0.2548$ &  $\frac{3}{2\sqrt{10}}$ & 
      $\frac{1}{\sqrt{24}}$ &  $\frac{1}{2\sqrt{15}}$ \\
$Q_R^e$ & $0.2316$ & $-0.3339$ &  $\frac{1}{2\sqrt{10}}$ & 
      $\frac{-1}{\sqrt{24}}$ & $\frac{2}{2\sqrt{15}}$ \\
$Q_L^\nu$ & $0.5$ & $0.2548$ &  $\frac{3}{2\sqrt{10}}$ & 
      $\frac{1}{\sqrt{24}}$ &  $\frac{1}{2\sqrt{15}}$ \\
\end{tabular}
\end{ruledtabular}
\end{table}

Before ending this section, we quote current limits on an extra $U(1)$ gauge
boson from direct searches at colliders.  The most stringent limits are given
by the preliminary results from CDF \cite{cdf-z} at the Tevatron:
\begin{eqnarray}
Z'_{\rm SM} & > & 845 \; {\rm GeV} \;, \nonumber \\
Z_{\chi} & > & 720 \; {\rm GeV} \;, \nonumber \\
Z_{\psi} & > & 690 \; {\rm GeV} \;, {\rm and} \nonumber \\
Z_{\eta} & > & 715 \; {\rm GeV} \;. \nonumber 
\end{eqnarray}
In the following, we will use a typical value of $M_{Z'}=1$ TeV unless
otherwise stated.

\section{Constraints from $D^0$-$\overline{D^0}$ mixing}

\subsection{$D^0$-$\overline{D^0}$ mixing in SM}

To second order in perturbation, the off-diagonal elements in the neutral $D$
meson mass matrix contain two contributions from short-distance physics.  One
part involves $|\Delta C| = 2$ local operators from box and dipenguin diagrams
\cite{Datta,Petrov:1997ch} at the $m_D$ scale, contributing to only the
dispersive part of the mass matrix.  Due to a severe CKM suppression in the SM,
the contribution from this part is negligible.  The other part involves the
insertion of two $|\Delta C| = 1$ transitions, contributing to both the
dispersive and absorptive parts of the mass matrix.

Since CP is a good approximate symmetry in $D$ decays, we have the CP
eigenstates $|D_{\pm}\ran$ with ${\cal CP}|D_{\pm}\ran = \pm |D_{\pm}\ran$ as
the mass eigenstates too.  It is convenient to define
\begin{eqnarray}
\Delta m_D = m_+ - m_- ~,
\qquad
\Delta\Gamma_D = \Gamma_+ - \Gamma_- ~,
\qquad
\overline\Gamma_D = \frac12 (\Gamma_+ + \Gamma_-) ~,
\end{eqnarray}
and consider the dimensionless parameters
\begin{eqnarray}
x_D \equiv \frac{\Delta M_D}{\overline\Gamma_D}
\qquad \mbox{and} \qquad
y_D \equiv \frac{\Delta\Gamma_D}{2\overline\Gamma_D} ~.
\end{eqnarray}

The short-distance contributions to $x_D$ and $y_D$ have been evaluated to the
next-to-leading order (NLO) and both found to be about $6 \times 10^{-7}$,
quoting the central values from Ref.~\cite{Golowich:2005pt}.  They are far
below the current experimental constraints.  In contrast, the long-distance
effects are expected to be more dominant but difficult to estimate accurately
\cite{Donoghue:1985hh}.

\subsection{$D^0$-$\overline{D^0}$ mixing in $Z'$ models}

As shown in Sec.~II, in $Z'$ models one can generate off-diagonal $Z'$ coupling
to charm and up quarks.  Due to the large $Z'$ mass, this can induce tree-level
processes for the $D^0$-$\overline{D^0}$ mixing.  Therefore, the $|\Delta C| =
2$ operators receive new contributions.  However, it has less influence on the
long-distance physics.  In view of the smallness of the SM contributions
through the double insertion of $|\Delta C| = 1$ operators, here we want to
estimate the pure $Z'$ effect on $x_D$, checking whether our model contradicts
with current experimental bounds on $D^0$-$\overline{D^0}$ mixing.

At the $M_W$ scale, the most general $|\Delta C| = 2$ effective Hamiltonian due
to the FCNC $Z'$ interactions is:
\begin{eqnarray}
{\cal H}_{\rm eff}^{Z'} & = & \frac{g_2^2}{2 M_{Z'}^2}
\left[ \bar{u} \gamma^\mu (C^{uc}_L P_L+ C^{uc}_R P_R) c \right]
\left[ \bar{u} \gamma_\mu (C^{uc}_L P_L+ C^{uc}_R P_R) c \right] + h.c. \;,
\end{eqnarray}
where $C_{L,R}^{uc}$ are generic left- and right-handed $Z'$ coupling to $u$
and $c$ quarks.  Since we suppose there is no flavor-changing couplings for the
right-handed fermions, $C^{uc}_R = 0$ and we obtain:
\begin{eqnarray}
{\cal H}_{\rm eff}^{Z'} & = & \frac{G_F}{\sqrt{2}} \left( \frac{g_2 M_Z }{g_1
  M_{Z'} } \right)^2 (C^{uc}_L)^2 {\cal O} + h.c. ~,
\end{eqnarray}
where $g_1=e/(s_W c_W)$ and ${\cal O} = [\bar{u}\gamma^\mu (1-\gamma_5)c]
[\bar{u} \gamma_\mu (1-\gamma_5) c]$.  Therefore, its contribution to the
neutral $D$ meson mass difference is
\begin{eqnarray}
\Delta m_D^{Z'}
& = & 2 \vert M_{12} \vert = 2 \frac{1}{2 m_D} 
<D^0\vert {\cal H}_{\rm eff}^{Z'} \vert \overline{D^0}> \nn \\
& = & \frac{8}{3} \frac{G_F}{\sqrt{2}} m_D f_D^2 B_D 
\left( \frac{g_2 M_Z}{g_1 M_{Z'} } \right)^2 (C^{uc}_L)^2 ~,
\label{deltamdp}
\end{eqnarray}
where $\lan D^0\vert {\cal O} \vert \overline{D^0} \ran = \frac{8}{3} m_D^2
f_D^2 B_D$ has been used.  Numerically, we obtain
\begin{eqnarray}
\Delta m_D^{Z'} &\simeq & 3 \times 10^{-8} B_D  
\left( \frac{1000 \,{\rm GeV} }{M_{Z'}} \right)^2 \left( C^{uc}_L \right)^2
\; \mbox{GeV} ~,
\end{eqnarray}
where we take $g_2 = g_1$ and the $D$ meson decay constant $f_D = 300$ MeV.  In
the vacuum insertion approximation, the bag parameter $B_D = 1$.  This is
translated into
\begin{equation}
x_D^{Z'} \simeq 2 \times 10^4 \left( C_{uc}^L \right)^2 ~.
\end{equation}
Note that $C_{uc}^L = Q^u_L (x-1) V_{ub} V^*_{cb} \simeq 1.5 \times 10^{-4}
(x-1) Q^u_L$, where we have neglected the renormalization group running effects
in comparison with the uncertainties in the $Q^u_L$ and the $x$ parameter in
the model.  Therefore, $x_D^{Z'} \simeq 4.6 \times 10^{-4} (x-1)^2 \left( Q^u_L
\right)^2$.

The current limits from the Dalitz plot analysis of $D_0 \to K_S \pi^+ \pi^-$
by CLEO are $(-4.5 < x_D < 9.3)$\% and $(-6.4 < y_D < 3.6)$\% at the 95\% C.L.
\cite{Asner:2005sz}.  Assuming negligible CP violation in the $D^0$ system, a
recent Belle analysis using the $D^0 \to K^+ \pi^-$ decays from $400$ fb$^{-1}$
of data yields to obtain $x_D^{\prime2} < 0.72 \times 10^{-3}$ and $-9.9 \times
10^{-3} < y'_D < 6.8 \times 10^{-3}$ at 95\% C.L. \cite{Zhang:2006dp}.  Here
the modified dimensionless parameters $x'_D = x_D \cos\delta_{K\pi} + y_D
\sin\delta_{K\pi}$ and $y'_D = y_D \cos\delta_{K\pi} - x_D \sin\delta_{K\pi}$,
with $\delta_{K\pi}$ being the strong phase difference between the
doubly-Cabibbo-suppressed and Cabibbo-favored amplitudes.  It is easy to see
that as long as the combination $(x-1) Q_L^u$ is less than about ${\cal O}(1)$,
the experimental bounds can be well satisfied in the various $Z'$ models that
we have mentioned in the previous Section.

\section{Collider Phenomenology}

\subsection{Decay width of $Z'$}

\begin{figure}
\centering
\includegraphics[width=5in]{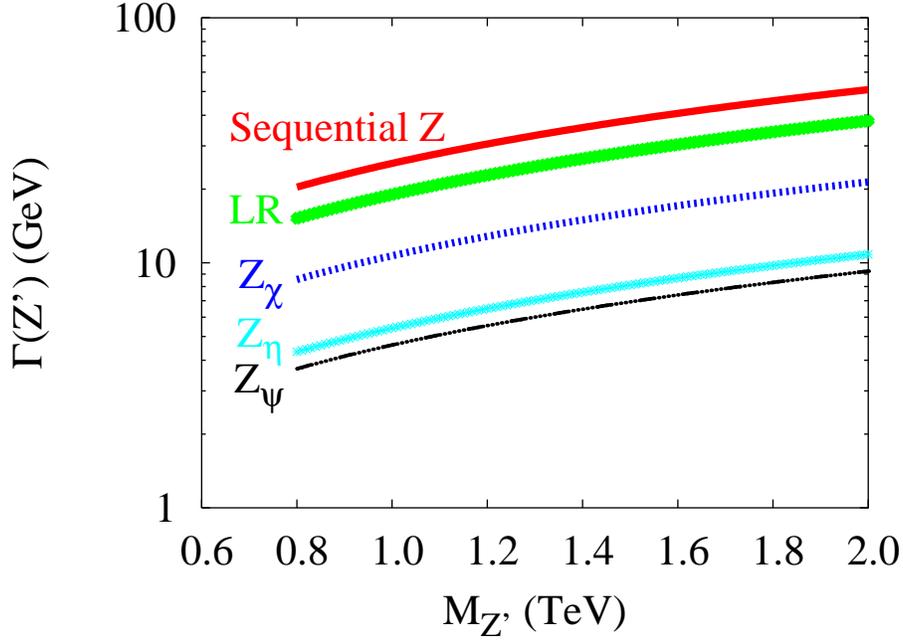}
\caption{\small \label{zwidth}
Total decay width of the $Z'$ boson in various $Z'$ models.}
\end{figure}

We include only the fermion modes in the computation of the $Z'$ decay width.
The partial decay width of $Z'\to W^+W^-$ is suppressed by the $Z$-$Z'$ mixing
angle which is severely constrained by electroweak precision data \cite{luo}.
Therefore, $Z'\to W^+W^-$ is not included in the total width.  The general
formula for the partial width into $f \bar f'$ is given by
\begin{eqnarray}
\Gamma (Z' \to f \bar f') &=& \frac{N_f g_2^2 M_{Z'} }{48 \pi}\, 
\lambda^{1/2}(1, \mu_1, \mu_2) \, \biggr [
\left( |Q_L^{ff'}|^2 + |Q_R^{ff'}|^2 \right ) \biggr (
   1 -\mu_1 - \mu_2  \nonumber \\
&& +(1+\mu_1 -\mu_2)(1-\mu_1 + \mu_2 ) \biggr )
 + 12 \sqrt{ \mu_1 \mu_2} \,\, {\cal R}e\, \left( Q_L^{ff'} Q_R^{ff'*} \right)
               \,  \biggr ]
\end{eqnarray}
where $N_f = 3(1)$ for quark (lepton) and $\lambda(1, \mu_1, \mu_2) = (1 -
\mu_1 - \mu_2)^2 - 4 \mu_1 \mu_2 $ with $\mu_1 = m_f^2/M_{Z'}^2$ and $\mu_2 =
m_{f'}^2/M_{Z'}^2$.  We include only the flavor diagonal modes $e^- e^+$,
$\mu^- \mu^+$, $\tau^- \tau^+$, $\nu_e \bar \nu_e$, $\nu_\mu \bar \nu_\mu$,
$\nu_\tau \bar \nu_\tau$, $u \bar u$, $d \bar d$, $c \bar c$, $s \bar s$, $t
\bar t$, and $b \bar b$ in the calculation.  Thus we have $Q_{L,R}^{ff'} =
Q_{L,R}^{f} \delta^{ff'}$, with $Q_{L,R}^e = Q_{L,R}^\mu = Q_{L,R}^\tau$,
$Q_{L}^{\nu_e} = Q_{L}^{\nu_\mu} = Q_{L}^{\nu_\tau} = Q_{L}^{\nu}$ , $Q_{L,R}^d
= Q_{L,R}^s = Q_{L,R}^b$, $Q_{L,R}^u = Q_{L,R}^c$ and $Q_{L,R}^t = x
Q_{L,R}^u$. The numerical values of these chiral couplings can be obtained from
Table~\ref{table1}.  We show the total decay width of the $Z'$ boson versus
$M_{Z'}$ in Fig.~\ref{zwidth}.  As it can be seen, the total width of $Z'$ is a
few to a few tens of GeV in all $Z'$ models that we list in
Table.~\ref{table1}.  It turns out that the branching fractions of all
fermionic modes are not sensitive to the $Z'$ mass.  For instance, the
branching fractions of $Z'\to q\bar q$, $Z'\to \nu\bar \nu$, $Z'\to
\ell^+\ell^-$ and $Z'\to t \bar t$ decays are in percentage of $68$, $20$, $10$
and $2$, respectively, for the sequential $Z$ model.  As we will see later,
even the largest flavor non-diagonal mode $t\bar c +\bar t c$ has a tiny
branching fraction of $10^{-3}$.  We will take a typical value of $M_{Z'} = 1$
TeV in subsequent analyses.  The decay widths of such a $Z'$ in various models
are give in Table~\ref{table2}.

\begin{table}[b!]
\caption{\small \label{table2}
Total decay widths of a $Z'$ of $M_{Z'} = 1$ TeV in various models.}
\medskip
\begin{ruledtabular}
\begin{tabular}{cccccc}
Model & Sequential $Z$ & $Z_{LR}$ & $Z_\chi$ & $Z_\psi$ & $Z_\eta$ \\
\hline
$\Gamma_{Z'}$ (GeV) & 27 & 20 & 11 & 4.9 & 5.7 \\
\end{tabular}
\end{ruledtabular}
\end{table}

\subsection{Hadronic production of $t \bar c + \bar t c$}

\begin{figure}[t!]
\centering
\vskip-3.2cm
\includegraphics[clip,width=4in]{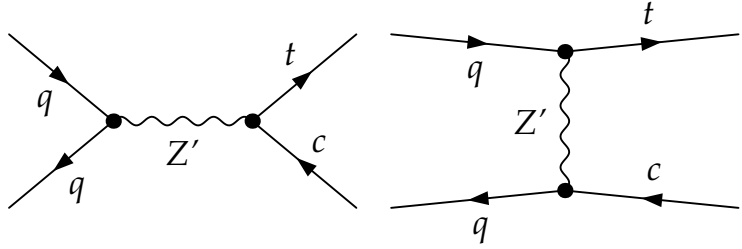}
\vskip-3.2cm
\caption{\label{fig1}
Contributing Feynman diagrams for 
$q\bar{q}$ annihilation into $t \bar c$ via the $Z'$ boson.}
\end{figure}

Let us define our notation for the convenience of the following formulas.  The
momenta of the incoming quark and anti-quark, outgoing top and outgoing
anti-charm quarks are denoted by $p_1$, $p_2$, $k_1$ and $k_2$, respectively.
We neglect the quark masses of the incoming partons.  The Mandelstam variables
are defined as follows
\begin{eqnarray}
\hat s &=&  (p_1+p_2)^2 = (k_1+k_2)^2  \nonumber \\
\hat t &=& (p_1-k_1)^2 = (p_2-k_2)^2 
= \frac{ m_t^2 + m_c^2}{2} - \frac{\hat s}{2} \left( 1 - \beta \cos\theta^*
                                              \right )
\nonumber \\
\hat u &=& (p_1-k_2)^2 = (p_2-k_1)^2 
=\frac{ m_t^2 + m_c^2}{2} - \frac{\hat s}{2} \left( 1 + \beta \cos\theta^*
                                              \right )
\nonumber \\
\hat u_c &=& \hat u - m_c^2\,, \qquad 
\hat u_t = \hat u - m_t^2\,,  \nonumber \\
\hat t_c &=& \hat t - m_c^2\,, \qquad 
\hat t_t = \hat t - m_t^2\,,  \nonumber \\
\hat s_{Z'} &=& \hat s - M^2_{Z'} + i \Gamma_{Z'} M_{Z'} \,, \qquad
\hat t_{Z'} = \hat t - M^2_{Z'} \nonumber 
\end{eqnarray}
where $\beta = \lambda^{1/2}(1, m_c^2/\hat s, m_t^2/\hat s)$ and $\theta^*$ is
the scattering angle in the center-of-mass frame of the partons.  The imaginary
part in the $\hat s_{Z'}$ is the Breit-Wigner prescription for regulating the
$Z'$ pole.

\begin{figure}[t!]
\centering
\includegraphics[width=5.9in]{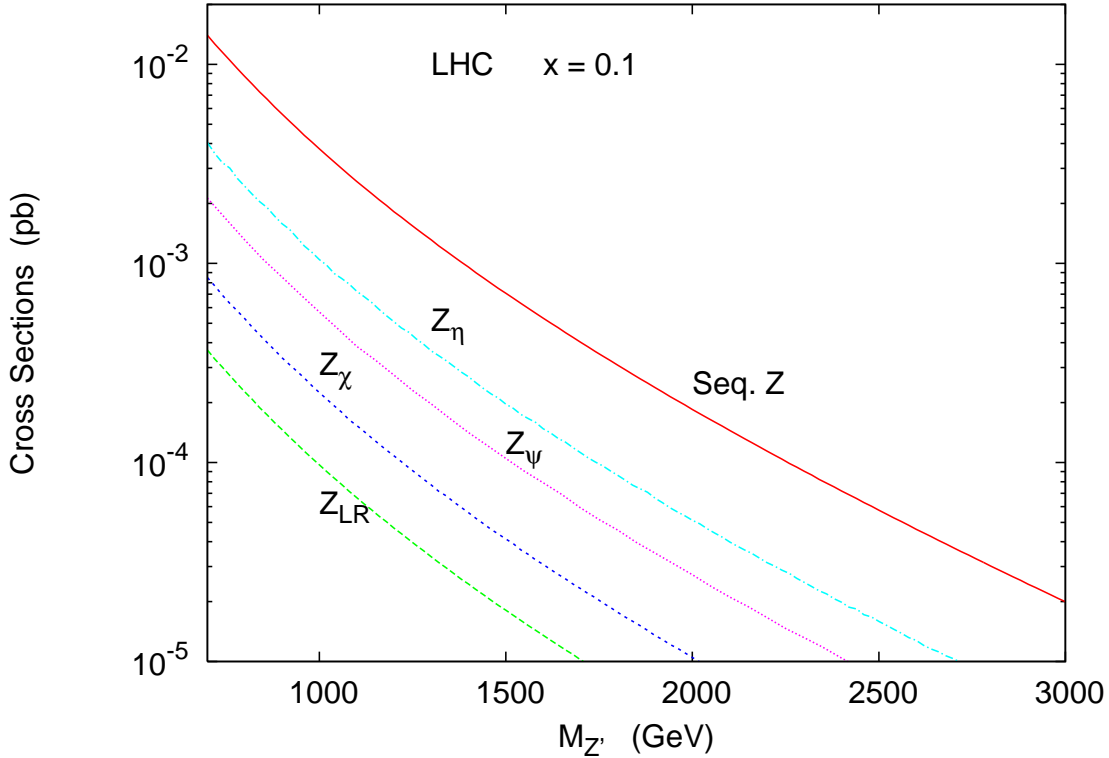}
\medskip
\caption{\small \label{z-x-section}
Production cross sections for $pp \to t\bar c + \bar t c$ at the LHC versus
the $Z'$ mass.  Results for the five models mentioned in the text
are presented.}
\end{figure}

At hadron colliders, the production proceeds via the conventional Drell-Yan
$s$-channel mechanism as well as the $t$-channel diagram (Fig.~\ref{fig1}).
The $s$-channel diagram dominates when $\sqrt{\hat s}$ is close to $M_{Z'}$.
Suppose we write the amplitude ${\cal M} = {\cal M}_s + {\cal M}_t$, after
summing over final-state helicities and colors and averaged over initial-state
helicities and colors, the amplitude squared is given by
\begin{eqnarray}
\overline{\sum} 
|{\cal M}|^2 &=&  \overline{\sum} |M_s|^2 + \overline{\sum} |M_t|^2
                 + \overline{\sum} \left ( M_s M_t^* + M_s^* M_t \right ) ~,
\end{eqnarray}
where
\begin{eqnarray}
\overline{\sum} 
|{\cal M}_s|^2 &=&  \frac{g_2^4}{ 4 \, \hat s_{Z'}^2} \, \Biggr [
 2 \left( |Q_L^q|^2 + |Q_R^q|^2 \right ) 
  \left( |Q_L^{tc}|^2 + |Q_R^{tc}|^2 \right ) ( \hat u_c \hat u_t 
                                       + \hat t_c \hat t_t ) \nonumber \\
&&
-2  \left( |Q_L^q|^2 - |Q_R^q|^2 \right ) 
   \left( |Q_L^{tc}|^2 - |Q_R^{tc}|^2 \right ) ( - \hat u_c \hat u_t + 
                                       \hat t_c \hat t_t )  \nonumber\\
&&
+ 8 m_c m_t \left( |Q_L^q|^2 + |Q_R^q|^2 \right ) \, {\cal R}e\, \left( 
  Q_L^{tc} Q_R^{tc*} \right) 
 \hat s   \Biggr ] ~,
\end{eqnarray}
\begin{eqnarray}
\overline{\sum} 
|{\cal M}_t|^2 &=& \frac{g_2^4}{ 4 \, \hat t_{Z'}^2} \, \Biggr [
  2 \left( |Q_L^{tq}|^2 + |Q_R^{tq}|^2 \right ) 
    \left( |Q_L^{qc}|^2 + |Q_R^{qc}|^2 \right ) \left( \hat u_c \hat u_t + 
                                 \hat s( \hat s-m_t^2 - m_c^2 ) 
\right. \nonumber \\ 
&& \left. \qquad     + \frac{m_c^2 m_t^2}{2 M_{Z'}^4} \hat t_t \hat t_c 
     + \frac{ 2 m_c^2 m_t^2}{ M_{Z'}^2} \hat s \right )
                                                            \nonumber \\
&&
- 2 \left( |Q_L^{tq}|^2 - |Q_R^{tq}|^2 \right ) 
    \left( |Q_L^{qc}|^2 - |Q_R^{qc}|^2 \right ) \left( -\hat u_c \hat u_t + 
                                \hat s( \hat s-m_t^2 -m_c^2)  \right) 
                                                      \Biggr ] ~,
\end{eqnarray}
and
\begin{eqnarray}
\overline{\sum} 
\left( {\cal M}_s {\cal M}_t^* \right. 
&+& \left. {\cal M}_s^* {\cal M}_t \right) \nonumber \\
&=& 
\frac{1}{3} \,
  \frac{ g_2^4}{ 2 \, \hat s_{Z'} \hat t_{Z'} } \, \Biggr [
   2 \,{\cal R}e\, \left( 
   Q_L^{tc} Q_L^{qc*} Q_L^{q} Q_L^{tq*} +
   Q_R^{tc} Q_R^{qc*} Q_R^{q} Q_R^{tq*}  \right ) \; 
   \left( 2 \hat u_c \hat u_t + \frac{m_c^2 m_t^2}{M_{Z'}^2} \hat s  \right )
   \nonumber \\
&&
 + 2 m_c m_t 
   \, {\cal R}e\, \left( 
   Q_L^{tc} Q_R^{qc*} Q_R^{q} Q_R^{tq*} +
   Q_R^{tc} Q_L^{qc*} Q_L^{q} Q_L^{tq*}  \right ) \;
   \left( 2 \hat s + \frac{ \hat t_c \hat t_t}{M_{Z'}^2} \right )\,\Biggr ] \;.
\end{eqnarray}
The interference term needs to be included for the subprocess $c \bar c \to t
\bar c$ only.  Thus, it simplifies down to
\begin{eqnarray}
\overline{\sum} 
\left( {\cal M}_s {\cal M}_t^* + {\cal M}_s^* {\cal M}_t \right) &=& 
 \frac{1}{3} \,
  \frac{g_2^4}{ 2 \hat s_{Z'} \hat t_{Z'} } \, \Biggr[
2 \, \left(   |Q_L^{tc}|^2 |Q_L^{c}|^2 +  |Q_R^{tc}|^2 |Q_R^{c}|^2 \right ) \;
\left( 2 \hat u_c \hat u_t + \frac{m_c^2 m_t^2}{M_{Z'}^2} \hat s  \right ) 
\nonumber \\
&&
+ 2 m_c m_t {\cal R}e\, \left( 
   Q_L^{tc} Q_R^{tc*} |Q_R^{c}|^2  +
   Q_R^{tc} Q_L^{tc*} |Q_L^{c}|^2    \right ) \,
   \left( 2 \hat s + \frac{ \hat t_c \hat t_t}{M_{Z'}^2} \right )\, \Biggr ] \,.
\end{eqnarray}
In the above equations, the chiral charges $Q_{L}^{tc}$, $Q_{L}^{tq}$ and
$Q_{L}^{qc}$ are given by the off-diagonal matrix elements of the matrix
$B^u_{L}$ defined in Eq.~(\ref{zput}).  For instance, $Q_L^{tc} = (x-1) Q_L^u
V_{cb} V_{tb}^*$ etc. Our simplified assumption made in Sec.~II implies that
all the off-diagonal right-handed chiral couplings vanish.
The partonic differential cross section in the parton rest frame is given by
\begin{equation}
\frac{d \hat \sigma}{d \cos \theta^*} =
 \frac{\beta}{32 \pi \hat s} \overline{\sum} |{\cal M}|^2 \;.
\end{equation}
The partonic cross section is then convoluted with the parton distribution
functions, for which the leading order fit (L) of the CTEQ6 sets \cite{cteq6}
are used.  We show the production cross section of $t\bar c + \bar t c$ at the
LHC versus the $Z'$ mass in Fig.~\ref{z-x-section}.  The major portion of the
cross section comes from the kinematic region where $\sqrt{\hat s}$ is close to
the $Z'$ mass.  We present our results only for $t\bar c + \bar t c$
production, it is clear from Eq.~(\ref{zput}) that the production rate for
$t\bar u + \bar t u$ is relatively suppressed by $|V_{ub}/V_{cb}|^2$.

\subsection{Detection of $t$ and $c$ and backgrounds}

As we have mentioned in the Introduction, the associated top-charm decay mode
of the $Z'$ can in principle be discovered via the Drell-Yan channel.  After
knowing the mass of the $Z'$ to some precision, one can look at the hadronic
decays of the $Z'$.  Among the hadronic decays containing the top-quark, we
determine if it contains one or two top-quarks.  If there are two top-quarks,
it may be just the flavor-diagonal decay of the $Z'$.  However, if there are
only one top plus an another heavy-flavor jet (the $c$ quark) in the hadronic
decays of the $Z'$, we identify it as the FCNC signal that we are searching
for.

The most serious irreducible background is the SM single top-quark production.
We calculate the SM single top-quark production cross section using MADGRAPH
\cite{mad}.  The single top-quark production receives contributions from the
following subprocesses
\begin{eqnarray}
&&  q \bar q' \to W^* \to t \bar b   + \bar t b \nonumber \\
&&  q g \to t\bar b j + \bar t b j \nonumber \\
&&  b g \to t j j \nonumber
\end{eqnarray}
where $j$ denotes a light quark jet.  It is mainly the $b$ quark in the final
state that may be mis-identified as the charmed jet of the signal.  Both the
charmed and bottom jets can be identified using the secondary vertex method,
and both of them can give rise to a displaced vertex in the silicon vertex
detector.  That is why the SM single top-quark production is the most serious
irreducible background in this FCNC signal search.  One can, however, use the
secondary vertex mass to further distinguish between the charmed and bottom
jets, as we shall explain in the next subsection.  Before we come to that, we
use some kinematic cuts to reduce the background cross section to about the
same level as the signal cross section \cite{Stelzer:1998ni}.

\begin{figure}[t!]
\centering
\includegraphics[width=6in]{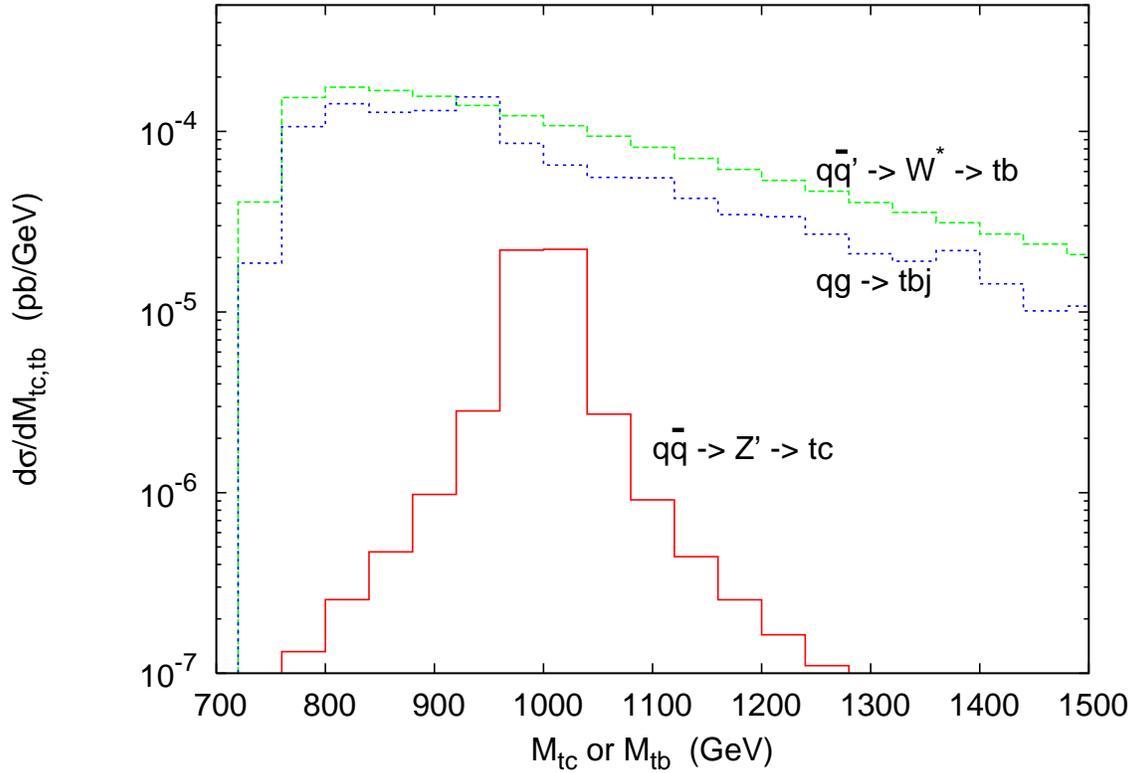}
\caption{\small \label{mtc}
Differential cross sections versus the invariant mass of the top-quark
and the heavy flavor (i.e. $M_{tc}$ or $M_{tb}$) for the sequential $Z$
model and the SM single top-quark backgrounds: 
$q\bar q' \to W^* \to tb$ and $qg \to tbj$ at the LHC.}
\end{figure}

One obvious cut to reduce this background is to require a very high transverse
momentum for the top-quark or the heavy-flavor jet, because of the heavy $Z'$.
We employ
\begin{equation}
\label{pt1}
p_T (t), p_T(j) \; > \; 350 \;{\rm GeV}\;, \qquad |y(t)|,|y(j)|<2.5
\end{equation}
for $M_{Z'}=1$ TeV.  The rapidity cut $|y(j)|<2.5$ is due to the coverage of
the central vertex detector.  The hadronic calorimetry can, however, go very
forward and backward up to about $y=4.5$ or $5$.  Since there is an additional
jet in the subprocess $qg \to t b j$, we can employ a jet veto to eliminate the
events with the third jet defined by
\begin{equation}
\label{pt2}
p_T(j) > 15 \;{\rm GeV}\;, \qquad |y(j)|<4.5  \qquad ({\rm Veto}) \;.
\end{equation}
In this way, the $qg \to t b j$ is reduced to a level smaller than $q\bar q'
\to tb$.  As we have explained above, we require to see only one heavy-flavor
jet with the top-quark.  Therefore, the subprocess $b g\to t jj$ is reduced to
a negligible level.  After imposing the cuts in Eqs.~(\ref{pt1}) and
(\ref{pt2}), we show in Fig.~\ref{mtc} the differential cross sections versus
the invariant mass of the top-quark and the heavy flavor ($c$ in the signal and
$b$ in the background).  In the figure, we illustrate the signal with the
sequential $Z$ model.  The total background is still about a factor of 5 larger
than the signal.  We have to rely upon the secondary vertex mass method or
$D$-, $D^*$-tagging to further separate the charmed and the bottom jets. We are
going to explain it in the next subsection.

\subsection{Charm Tagging}

Heavy quark flavor tagging is in general quite successful up to some
limitation. We briefly describe it here.  With the silicon vertex detector, one
can use the presence of a secondary vertex in a jet to identify it as a
heavy-flavor jet.  The presence of a secondary vertex in a silicon vertex
detector is in general due to the long decays of a bottom or charmed hadron.
Here one requires at least two tracks (the minimum to form a secondary vertex)
to meet at a point far away enough from the interaction point.  A positive tag
is placed when the secondary vertex is more than two standard deviations from
the interaction point.

Once a jet is identified with as a heavy-flavor jet, one can measure the
secondary vertex mass (the invariant mass of the hadrons at the secondary
vertex) to further distinguish between the charmed and bottom jets.  A
distinctive figure shown in Ref.~\cite{charm-tag} clearly shows the difference
among the charmed, bottom, and $uds$-jets.  The bottom jet has the largest
secondary vertex mass with a tail up to 4 GeV, while the charmed jet has a
secondary vertex mass ranging from 0 to 2 GeV with a peak around 1 GeV.  The
light quark jets have the smallest secondary vertex masses.  One can make use
of the Monte-Carlo templates to determine the fractions of charm, bottom, and
other light quarks in a jet sample.

Another method is to identify the $D^*$ and $D$ mesons, which the prompt
charm-quark hadronizes into.  It has been used to measure the prompt charmed
mesons production at the Tevatron \cite{cdf-charm}.  One can reconstruct the
charmed mesons in the following decay modes:
$D^0 \to K^- \pi^+$, $D^{*+} \to D^0 \pi^+$ with
$D^0 \to K^- \pi^+$, $D^+ \to K^- \pi^+ \pi^+$,
$D_s^+ \to \phi \pi^+$ with $\phi \to K^+ K^-$,
and their charge conjugates.  Details of reconstructing these charmed mesons
can be found in Ref.~\cite{cdf-charm}.  The most important criterion is to
distinguish between the prompt charmed mesons and those from bottom meson
decays.  These two sources can be separated using the impact parameter of the
net momentum vector of the charm candidate to the beamline.  Prompt charmed
mesons will point back to the beamline because the charm-quark hadronizes
immediately after it is produced.  Therefore, one can have some success in
tagging the prompt charmed meson together with a single top-quark.  However, we
anticipate the efficiency not to be too high.  The realistic efficiency is
beyond the scope of the present paper.

We summarize our findings for the LHC as follows.
\begin{enumerate}
\item The FCNC production of $t\bar c + \bar t c$ is mainly via on-shell
  production of the $Z'$ boson.  The most likely scenario is that the $Z'$
  boson is first discovered in the gold-plated channel, the Drell-Yan process.
  We then search for the production of a single top-quark and a charmed jet in
  the hadronic decay of the $Z'$ boson.  Since the single top-quark and the
  charmed jet originate from the $Z'$ decay, we impose a very large $p_T$ cut
  to reduce the background.  The production rate of $t\bar c + \bar t c$ for
  $M_{Z'}=1$ TeV is of the order of $1$ fb for several typical $Z'$ models that
  we study in this work.
\item The most serious irreducible background is the SM single top-quark
  production associated with a bottom-quark.  The collider signatures for a
  charmed jet and for a bottom jet are similar.  Both have a secondary vertex
  in the silicon vertex detector.  One may be able to use the secondary vertex
  mass method or to use the $D,D^*$-meson tagging to distinguish between the
  charmed and bottom jets.  However, experimental separation of charmed and
  bottom jets is still uncertain, so one would expect some difficulty in
  getting a clean signal.  One has to rely on an accurate estimation of the SM
  background in order to extract the signal.
\end{enumerate}

\subsection{$e^+e^-\to t \bar c \, + \, \bar t c$ at ILC}

\begin{figure}[t!]
\centering
\includegraphics[width=6in]{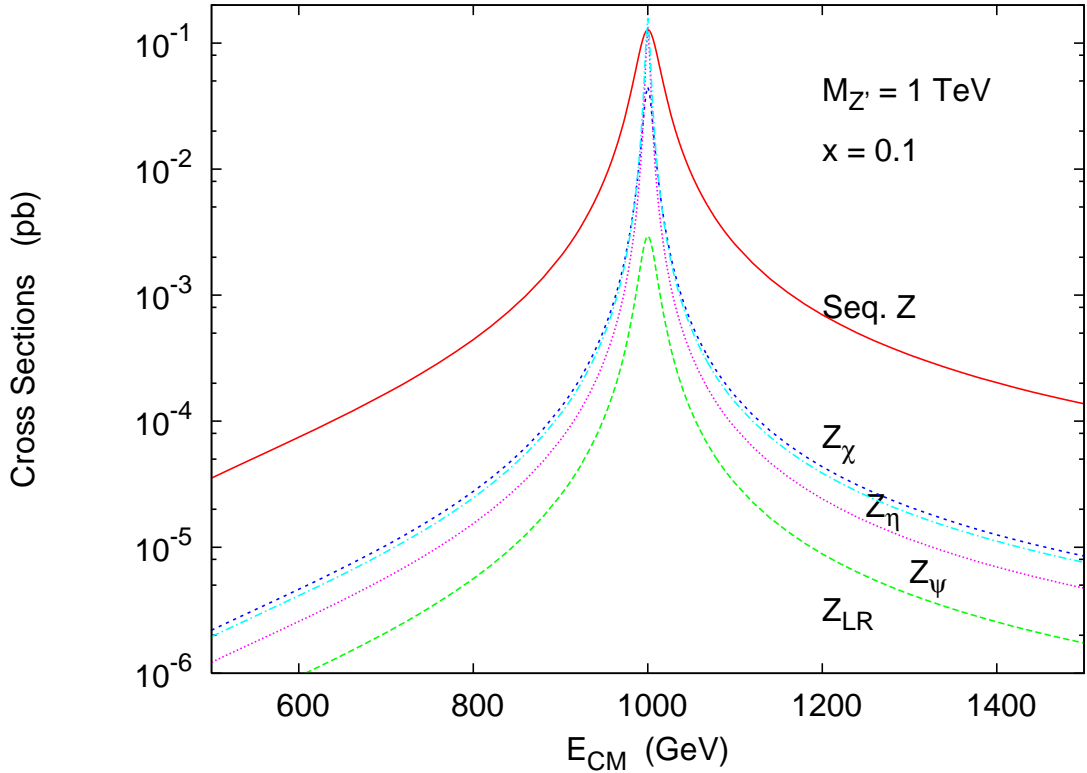}
\medskip
\caption{\small \label{ee-x-section}
  Production cross sections for $e^+ e^- \to t\bar c + \bar t c$ at a linear
  $e^+ e^-$ collider versus the center-of-mass energy.  Results for the five
  models mentioned in the text are presented.}
\end{figure}

At linear colliders such as the ILC, only the $s$-channel diagram contributes
to the process $e^+e^-\to \bar{t}c$ or $t \bar c$.  The differential cross
section can be adapted from the above formulas and it reads
\begin{eqnarray}
\frac{d \sigma}{d \cos \theta} &=&
\frac{3 g_2^4 \beta}{64 \pi s} 
\frac{1}{ s_{Z'}^2} \, \Biggr [
  \left( |Q_L^e|^2 + |Q_R^e|^2 \right ) 
  \left( |Q_L^{tc}|^2 + |Q_R^{tc}|^2 \right ) ( u_c  u_t 
                                       + t_c t_t ) \nonumber \\
&& -\left( |Q_L^e|^2 - |Q_R^e|^2 \right ) 
   \left( |Q_L^{tc}|^2 - |Q_R^{tc}|^2 \right ) ( -  u_c u_t + 
                                       t_c t_t )  \nonumber\\
&&
+ 4 m_c m_t \left( |Q_L^e|^2 + |Q_R^e|^2 \right ) \, {\cal R}e\, \left( 
  Q_L^{tc} Q_R^{tc*} \right) 
 s   \Biggr ] \;.
\end{eqnarray}
We show in Fig.~\ref{ee-x-section} the cross sections of $t\bar c + \bar t c$
production for $\sqrt{s}=0.5$ to 1.5 TeV with a fixed $Z'$ mass of 1 TeV.

\begin{figure}[t!]
\centering
\includegraphics[width=6in]{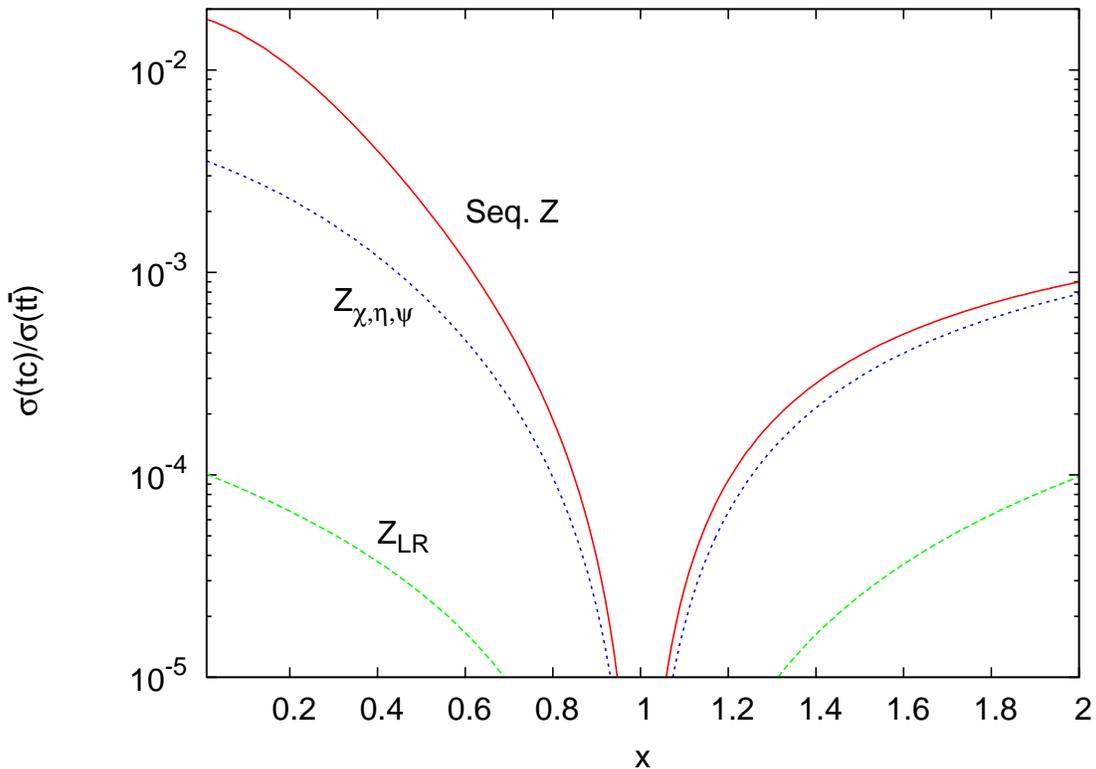}
\medskip
\caption{\small \label{ratio}
The ratio $\sigma(t\bar c+\bar t c)/\sigma(t\bar t)$ versus $x$ with
$\sqrt{s}=M_{Z'}=1$ TeV for
the models of sequential $Z$, $Z_{LR}$, $Z_\chi$, $Z_\eta$, and $Z_\psi$.
Note that the curves for $Z_\chi$, $Z_\eta$, and $Z_\psi$ overlap because
the ratios of the left-handed to the right-handed couplings are the 
same for these three $Z'$ models.}
\end{figure}

Unlike the case of the LHC, the detection of $t\bar c+ \bar tc$ at an $e^+ e^-$
collider is much more straight-forward because the SM single top-quark
production proceeds through $\gamma$-$t$-$q$ and $Z$-$t$-$q$ FCNC couplings
$(q=u,c)$ that are GIM suppressed \cite{tcsm}.  One can measure under the $Z'$
peak the cross sections for $t\bar t$ and $t\bar c +\bar t c$, and thus
determine the parameter $x$.  In fact, the ILC \cite{ilc} may have the option
of tuning the center-of-mass energy of the collision.  Then one can tune it to
the $Z'$ mass to maximize the production cross section, as shown in
Fig.~\ref{ee-x-section}.  With the silicon vertex detector one can detect
events with a heavy flavor (the charm-quark) and a single top-quark.  We show
in Fig.~\ref{ratio} the ratio of $\sigma (t\bar c + \bar t c)/ \sigma(t\bar t)$
versus the parameter $x$ for the $Z'$ models that we consider in this paper.
For a reasonable range of $x$ the ratio is about $10^{-4} - 10^{-2}$. Moreover,
the number of FCNC events under the $Z'$ peak is large, of the order of $10^3 -
10^4$ events for an integrated luminosity of $100$ fb$^{-1}$.  Therefore, we
conclude that the ILC will be much better than the LHC in probing the FCNC
process of $t\bar c + \bar t c$ production via a $Z'$ boson.

\section{conclusion}

In the framework of models with an extra $U(1)$ gauge boson that has family
non-universal couplings, one can induce tree-level FCNC couplings in the
up-type and/or down-type quark sector after diagonalizing their mass matrices.
In the models considered in this paper, it is possible to have tree-level
$Z'$-$t$-$c$ and $Z'$-$c$-$u$ couplings.  We have studied the collider
signature of associate top-charm production at both LHC and ILC and discussed
the constraint of a tree-level $Z'$-$c$-$u$ coupling from $D-\overline{D}$
mixing.

For the LHC, the main contribution to the associate $t\bar c + \bar t c$
production is from the $s$-channel diagram $qq\to Z^{\prime *}\to t\bar c +
\bar t c$. The total cross section can be of the order of $1$ fb for $M_{Z'}=1$
TeV in the framework of $Z'$ models that we have discussed.  The most serious
irreducible background is the SM single top-quark production associated with a
bottom-quark.  It is found that the total SM background is larger than the
signal. Therefore, in order to extract the FCNC signal of the $Z'$ boson, a
detailed Monte-Carlo simulation study of the SM background is required.

At the ILC, the situation is more promising given the fact that the signal is
almost background free. For $\sqrt{s}\approx M_{Z'}$, the cross section can
reach a size of more than 100 fb. Away from the resonance, there is still a
region where the cross section can be larger than the threshold of
observability $0.01$ fb for such a clean process.

\section*{ACKNOWLEDGMENTS}

This research was supported in part by the National Science Council of Taiwan
R.~O.~C.\ under Grant Nos.\ NSC 94-2112-M-007-010- and NSC 94-2112-M-008-023-,
and by the National Center for Theoretical Sciences.


\end{document}